\documentstyle[aps,prl,twocolumn,psfig]{revtex}
\begin{document}
\draft

\title{A non extensive approach to the entropy of symbolic sequences}
\author{Marco Buiatti $^{1}$\footnotemark[1], Paolo Grigolini$^{1, 2,
3}$\footnotemark[2], Luigi
Palatella$^{3}$\footnotemark[3]}

\address{$^{1}$Istituto di Biofisica del Consiglio Nazionale delle
Ricerche, Via San Lorenzo 26, 56127 Pisa, Italy }
\address{$^{2}$Center for Nonlinear Science, University of North
Texas, P.O. Box 305370, Denton, Texas 76203 }
\address{$^{3}$Dipartimento di Fisica dell'Universit\`{a} di Pisa,
Piazza
Torricelli 2, 56127 Pisa, Italy } \date{\today}
\maketitle
\footnotetext[1]{*e-mail: marcob@ib.pi.cnr.it}
\footnotetext[2]{\dag e-mail: grigo@soliton.phys.unt.edu}
\footnotetext[3]{\ddag e-mail: palatell@mann2.difi.unipi.it}

\begin{abstract}
Symbolic sequences with long-range correlations are expected to
result in a slow regression to a steady state of entropy increase.
However, we prove that also in this case a fast transition
to a constant rate of entropy increase can be obtained,
provided that the extensive entropy of Tsallis with entropic index $q$
is adopted, thereby resulting in a new form of entropy that
we shall refer to as Kolmogorov-Sinai-Tsallis (KST) entropy.
We assume that the same symbols, either $1$ or $-1$, are repeated in
strings
of length $l$, with the probability distribution $p(l)\propto
\frac{1}{l^{\mu}}$. The numerical evaluation of the KST entropy
suggests that at the
value $\mu = 2$ a sort of abrupt
transition might occur. For the values of $\mu$ in the range $1<\mu<2$
the entropic index $q$ is expected to vanish, as a consequence of
the fact that in this case the average length $<l>$ diverges, thereby
breaking the balance between determinism and randomness in favor
of determinism.  In the
region $\mu \geq  2$ the
entropic index $q$ seems to depend on $\mu$ through the
power law expression $q = (\mu-2)^{\alpha}$ with $\alpha\approx 0.13$
($q = 1$ with $\mu > 3$).
It is argued that this phase-transition like property signals the onset
of the thermodynamical regime at $\mu = 2$.
\\
\end{abstract}

        It has been recently pointed out\cite{ANISHCHENKO}
that power law spectra are observed in many disciplines
of science ranging from astronomy, geography and physics to
electronics, acoustic, linguistic and music. It is also
 interesting to establish a connection
between these observed properties and their algorithmic complexity.
This is important not only from a conceptual
point of view\cite{EBELINGALONE}: It also
might result in methods for the detection itself of correlations. In
this respect, we want to mention the search for correlations in DNA
sequences based on the adoption of entropic
indicators\cite{EBELINGFEISTEL,LI,FURTHER1,FURTHER2,FURTHER3}.

It has been remarked \cite{GASPARD}, however, that something
intermediate
between periodic and chaotic dynamical behavior exists and that
suitable tools to analize these processes must be built up.
These conclusions are widely shared in literature.
For instance, also the authors of
Refs.\cite{ANISHCHENKO,ebeling,ebeling1} as well as those of
Ref.\cite{GASPARD}, show that the entropy
of symbolic sequences in the case of long-range correlations
exhibits a regression to the condition of
constant Kolmogorov entropy
which turns out to be very slow. Analogous results are found
in many other papers\cite{WANG,FREUND,FREUNDEBELING} as well as
in earlier papers\cite{SZEPFALUSY}.

We shall refer ourselves to the Kolmogorov entropy applied to the
symbolic
sequences as metric
entropy (ME) \cite{BECK} to
keep it distinct from the Kolmogorov-Sinai entropy (KSE)
\cite{Kolmogorov,Sinai}. The two entropies are closely related to one
another, since both entropies are expressed in terms of the Shannon-Gibbs
entropy.
However,
the latter, the KSE, refers to individual trajectories and, in
principle, does not imply any coarse-graining if the assumption is
made that
cells and time steps of arbitrarily small size can be used.
The former applies to symbolic sequences
and consequently might be affected by a so large coarse-graining
process as to lose a direct connection with the rules,
either stochastic or deterministic, from which the sequence is
generated. This aspect will be made more transparent
by the discussion of the numerical experiment described in this paper.

The main purpose of this paper is that of discussing the
consequences of expressing the ME in terms of the
Tsallis entropy \cite{CONSTANTINO88}
rather than of the Shannon entropy. This is a form of ME that we
shall refer to as Kolmogorov-Sinai-Tsallis (KST) entropy. The Tsallis
 entropy reads
 \begin{equation}
 H_{q}= \frac{1 - \sum\limits_{i = 1}^{W}p_{i}^{q}}{q-1}.
 \label{entropy}
 \end{equation}
 Note that this entropy is characterized by the index $q$ whose
 departure from the
 conventional value $q = 1$ signals the thermodynamic effects of
 either long-range correlations in fractal dynamics or the non-local
 character of quantum mechanics\cite{luigi}. The increasing
  interest for
 Tsallis' non-extensive entropy
 is testified by the exponentially
 growing list of publications on this hot issue\cite{list}.

 Of remarkable interest for the subject of fractal dynamics is the
 discovery recently made by Tsallis \emph {et al.}\cite{TPZ97} that
 the entropic index $q$ also determines the specific analytical
 form illustrating the trajectory instability. Two trajectories,
 moving from infinitelly close but distinct initial conditions,
 depart from one another with a law more general than the exponential
 prescription. The exponential instability is a sort of singularity,
 namely,
 a special case of a more general, non-exponential, prescription. This
 important result is based
 on the generalization of the KSE \cite{Kolmogorov,Sinai}
 and consequently of the theorem of Pesin \cite{pesin}.
 Palatella and Grigolini\cite{luigi} have recently corroborated the
 conclusions of Miller and Sarkar\cite{miller&sarkar} who prove
 that in the quantum case the Von Neumann entropy is
 linearly proportional to the KSE. Furthermore, the results of
 these authors have been extended \cite{luigi} to the case where the quantum
 expression for the entropy (the von Neumann entropy) is expressed in
 terms of the Tsallis prescription. The interesting conclusion is that
 $q < 1$: Palatella and
 Grigolini\cite{luigi} argue that this result is a
 reflection of the occurrence of the Anderson localization.

 The present paper is devoted to discussing the convenience
 of the KST entropy to reveal whether or not a symbolic
 sequence does have or not a
 thermodynamical nature. The discussion rests on a key experiment,
 planned for the specific purpose of establishing correlations in
 sequences of symbols. The sequence of symbols is established as
 follows. Two computer generators of random numbers, $x$ and
 $z$, are used. The former generates random numbers distributed with
 equal probability in the interval $[0,1]$ and the latter is the
 generator of the fluctuations $z =+1$ and $z= -1$, with the same
 statistical weight. The uncertainty associated with each drawing of
 the numbers $x$ is
 \begin{equation}
h_{x} = lnW_{x},
\label{uncertainty1}
\end{equation}
where $W_{x}= 1/\Delta_{x}$ and $\Delta_{x}$ denotes the resolution
of the former random generator.
The drawing of the numbers z is equivalent to tossing a coin, and
consequently is associated with the uncertainty
 \begin{equation}
h_{z} = ln2.
\label{uncertainty2}
\end{equation}

Let us immagine now that at regular intervals of time, with the time
step $\Delta t = 1$, we draw a number $x$ and a number $z$. The
uncertainty $H(N)$ grows as a linear function of the number of
drawings $N$,
 \begin{equation}
H(N) = N(ln2 +lnW_{x}).
\label{uncertaintyN}
\end{equation}
We introduce a deterministic rule into this
totally stochastic picture.
This is done by replacing the variable $x$ with the
variable $y$ related to $x$ by
\begin{equation}
y = A [\frac{1}{(1-x)^{\frac{1}{\mu - 1}}} -1].
\label{yfromx}
\end{equation}
The probability distribution of the variable $y$ is given by
\begin{equation}
p(y) = (\mu-1) \frac{A^{\mu - 1}}{(A + y)^{\mu}}.
\label{probabilitydistribution}
\end{equation}
Note that the first moment of the variable $y$ is given by
\begin{equation}
<y> =  \frac{A}{\mu - 2}.
\label{firstmoment}
\end{equation}
This means that the value $\mu = 2$ is a critical point at which the
first moment of the new variable $y$ diverges.

The symbolic sequence is obtained by drawing the number $x$ first.
This number determines the number $y$ according to Eq. (\ref{yfromx}), and
fixes the number of sites $N_{y}= [y] + 1$, with $[y]$ denoting the
integer part of $y$, to fill with the same symbol (either $1$ or
$-1$). Then
we draw the number $z$, and we fill these sites either with $+1$ or
with $-1$ according to whether we get $z=1$ or $z=-1$.
Note that, according to the treatment of \cite{ALLEGRO}, the length of the
strings with the same symbols
(either $+1$
or $-1$) is proportional to the length of the laminar
regions generated by the nonlinear
maps that are currently used to mimic turbulent phenomena\cite{klafter}.
For this reason we shall refer to them as \emph{laminar strings}.
We can thus provide further support to our conviction
that critical properties have to be expected
at
$\mu = 2$. In fact we notice that the
divergence of Eq. (\ref{firstmoment})
at $\mu = 2$ implies that the
mean length of the laminar strings is infinite, and that,
consequently, once one symbol is known, the chances of guessing correctly a large number of symbols coming afterwards are high. Perhaps, a more proper way of illustrating
the region with $1< \mu \leq 2$, where all the moments of the
distribution $p(y)$ of Eq. (\ref{probabilitydistribution}) diverge,
 is that of referring to it as the
region where the balance between randomness and determinism is broken
and determinism prevails \cite{nota}.
 In conclusion, we think that the deterministic nature
 of this region can be properly denoted by the entropic index $q = 0$.
 On the other hand, we expect that the region $\mu>3$ is characterized
 by $q = 1$. This is so because the region $\mu>3$ implies that the
 second moment, as well as the first moment of $p(y)$, is finite,
 thereby ensuring the validity of the central limit theorem, and with
 it, of ordinary statistical mechanics. This means that $\mu>3$ is
 expected to yield $q = 1$.

The original uncertainty of Eq. (\ref{uncertaintyN}) is
deformed by the nonlinear transformation of Eq. (\ref{yfromx}).
However, as we have seen, this can force $q$ to depart from the
usual statistical value $q = 1$ only in the region $\mu<3$. The region
$\mu<2$ is expected to yield $q = 0$. We are thus only left with
the problem of establishing the dependence of $q$ on $\mu$ in the
region $2<\mu \leq 3$.  This can be done evaluating numerically
the KST entropy as follows.
After defining a given sequence, we fix
a window of length
$N$. Note that this length $N$ from now on will be referred to as {\em time}.
Then we move this window along the sequence generated according
to the rules earlier illustrated. For any position of this window we
find a given configuration
$A_{1} \bullet A_{2} \bullet... \bullet A_{N}$,
where
the $A_{i}$'s have either the value  $+1$  or the value $-1$.
We count the number of
configurations of the same kind
obtained moving the window along the chain, then we divide the number
of
these configurations by the total number of possible configurations
$W(N)$, thereby determining the probabilities $p_{i}$. Finally we use
Eq. (\ref{entropy}) to evaluate the entropy corresponding to this
window
of length $N$. It is convenient to study all this in the specific
case where the symbolic sequence is generated with no correlation among
the distinct sites. In this specific case $W(N) = 2^N$ and, of course,
$p_{i}(N) = 1/2^{N}$. Thus we obtain
\begin{equation}
H_{q}(N) = \frac{1 - 2^{N(1-q)}}{q-1}.
\label{nocorrelation}
\end{equation}

 Fig.\ref{fig1} illustrates the behavior
of $H_{q}(N)$ of Eq. (\ref{nocorrelation}) for different values of the
entropic index $q$. The middle line denotes the behavior corresponding
to $q = 1$. Let us identify, therefore, $q = 1$ with
$q_{true}$, namely, the entropic index properly reflecting a given
statistical
condition, the total absence of correlations, in this case.
Then we see that for $q < q_{true}$ the time derivative of entropy
tends to increase upon increase of
the time $N$. We
see
also that if the probing index $q$ is
larger than the correct entropic index, namely, $q >q_{true}$, the time
evolution of $H_{q}(N)$ is characterized by a rate of increase smaller
than the increase linear in time.
It is plausible that
the same qualitative behavior is present even if $q_{true}\neq 1$.
In fact, if we assume this behavior to be valid in general, namely,
even in the case where $q_{true}< 1$, we predict that the entropy growth
for $q =1$ is slower than that of a linear function of time, thus fitting
the observation made by several
authors (see, for instance, the
work of Ref. \cite{EBELINGALONE}).
This is an important remark since our main purpose here is to
apply our statistical analysis
to the case where the symbolic sequence is characterized by extended
correlations, and consequently the entropic index is expected
to depart
from the normal value $q_{true}= 1$.

However,
before addressing this challenging problem, it
is convenient to recall some important properties. First of all,
it is
worth noticing that the rules earlier adopted are equivalent to those
used in recent papers \cite{DNA1,DNA2,DNA3,DNA4} to build up
sequences that
turn out to be
statistically equivalent to the real DNA sequences.
The distributions
of $+1$, corresponding to purines, and of $-1$, corresponding to
pirimidines, was actually established adopting a nonlinear
map \cite{nonlinearmap}.
The nonlinear map adopted, in turn, was the same as that widely used
in the recent few years to generate anomalous diffusion. In a more
recent paper \cite{ALLEGRO}, it has been shown that these
nonlinear maps produce
effects statistically equivalent to a stochastic generator which is,
in fact, the same as that earlier illustrated as a generator of
long-range correlations in the symbolic sequences under study in this
paper.

Let us focus now our attention on the fact that any finite string
 $A_{1} \bullet A_{2} \bullet... \bullet A_{N}$ can be associated
 to an erratic trajectory moving from the ``time'' $i = 1$ to the ``time''
 $i = N$ on an one-dimensional lattice. The correspondence is
 established using the following prescription. At the time i the
 random walker makes a jump of unit length to the right or to the
 left according to whether $A_{i}= 1$ or $A_{i}= -1$. It is
 shown\cite{ALLEGRO} that in the case of a random walker with
 correlations infinitely extended in time there is a significant
 probability that the random walker might make N steps in the same
 direction. Thus, in a process of diffusion, with all the walkers
 initially concentrated in the same site, the distribution will split
 into two ballistic peaks moving in opposite directions. With the
 increase of N an increasing number of walkers belonging to a peak
 moving in a given direction will make jumps in the opposite direction.
 Thus the intensity of the side peaks of the distribution is a
 decreasing function of time, known\cite{ALLEGRO} to be proportional
 to the correlation function
 $\Phi(k) \equiv <A_{i}A_{i+k}>/<A_{i}A_{i}>$. It is evident that the
 strings $A_{1} \bullet A_{2} \bullet... \bullet A_{N}$ with all the
 $A_{i}'s$ equal to either $+1$ or $-1$, have the same intensity as
 these
 side peaks, to which these strings are equivalent.
 For this reason we shall refer ourselves to these side strings,
 with the same length as that of the exploring window of size $N$
 and with all the symbols $A_{i}$ corresponding to the same letter, as
 $\emph{border strings}$.

 For the sake of some preliminary remarks we make the simplifying assumption that all the strings  but the border
 strings have the same probability  $p(N)$. The
 dependence of $p(N)$ on $N$  is established by setting the
 normalization condition which yields
 \begin{equation}
p(N) = \frac{1 - 2\Pi(N)}{2^{N}-2}.
\label{centralpart}
\end{equation}
As earlier remarked, according to Ref.\cite{ALLEGRO}
\begin{equation}
\Pi(N) = \frac{\Phi(N)}{2}.
\label{peakintensity}
\end{equation}
Thus, under the assumption of equal probability for all the strings but the
border strings, we can write
\begin{equation}
H_{q}(N) = \frac{2\Pi(N)^{q}+(2^{N}-2)^{1-q}(1 - 2 \Pi(N))^{q}- 1} {1 - q},
\label{entropydominatedbybordersequences}
\end{equation}
with $\Pi(N)$ given by Eq. (\ref{peakintensity}). We note that, in
principle, the rules adopted to establish long-range correlations in
the sequences under study in this paper, make the correlation function
$\Phi(N)$ read
\begin{equation}
\Phi(N)  = \frac{A^{\beta}} {(A + N)^{\beta}},
\label{correlationfunction}
\end{equation}
where $\beta = \mu -2$. Consequently, the decay of these border
strings is extremely slow and it dominates the entropy time evolution
for a long time. On the other hand, for times so long as to make the
contribution of the central part more important than that resulting
from the border strings, the correct entropic index is given
by $q = 1$,
in accordance to the fact that in such a condition the statistical
properties of the sequences become indistinguishable from that
of totally uncorrelated sequences\cite{nota}. This is in line with the fact that
the diffusion process\cite{ALLEGRO} resulting from these rules is
characterized by two distinct rescaling properties, the ballistic
rescaling of the peaks and the L\'{e}vy rescaling of the central part of
the diffusion. This means that the interesting statistical properties
are blurred by the presence of the border strings. For this reason,
we decided to disregard the border strings and to set the
normalization condition only on the other strings.

In principle, if no length limitation were set on the
analysis of data, it would be
possible to derive the correct statistical properties by examining
suitably large windows. However,
for the sake of computational simplicity
we set the maximum length of the window to be $N_{max}= 10$.
On the other hand, as we shall see, the approach based on
disregarding the border strings makes it
possible to reveal the effect of correlations on the entropic index
with sequences of relatively small length.

 It has to be stressed that
the detection of the proper entropic index becomes more and
more difficult as the power index $\mu$ comes
closer and closer to the critical value $\mu = 2$. In fact,
the probabilities of given strings of length $N$ are closely related
to the correlation functions. The correlation function $\Phi(N)$,
for instance,
 on the basis of
the Shannon-McMillan-Breiman
theorem \cite{theorem1,theorem2,theorem3},
is the probability of a string of length equal to $2$. This makes it
possible
to explain why the finite length of the sequence yields
an error on the numerical
evaluation of the probability of a given sequence and
suggests how to correct this error. In fact, the finite
length of the sequence causes the truncation of the longer strings and,
consequently, the decay of correlation function $\Phi(N)$
becomes faster than theoretically expected on the
basis of the prescription of Eq.
(\ref{probabilitydistribution}). Thus, rather
than expressing the entropic index $q$ in terms
of the parameter $\mu$ corresponding to the prescription
of Eq. (\ref{probabilitydistribution}), we relate
$q$ to an effective $\tilde{\mu}$, obtained from the
numerical evaluation of the correlation function $\Phi(N)$.
More precisely, we determine numerically the
parameter $\beta$ and from it $\tilde{\mu} = 2 + \beta$.
The numerical results show that for values of
$\tilde{\mu} \approx 2.3$ or larger, the effective power index
coincides with the value that theoretically
should correspond to Eq. (\ref{probabilitydistribution}).

These numerical expedients make it possible for us
to bring the determination of $q$
as a function of
$\mu$, much closer to the critical region
$\mu = 2$. The method adopted is illustrated by Fig.\ref{fig2}. As
expected, on the basis of the results illustrated by Fig.\ref{fig1} 
and concerning the theoretical prescription
of Eq. (\ref{nocorrelation}), there exists a crucial value
of $q$,  which results in a linear dependence
of $H_{q}(N)$ on $N$. In Fig.\ref{fig2}, for instance, we see that
at $\tilde{\mu} \cong \mu = 2.5$ the solid line, corresponding
to $q \approx 0.89$ fits very well a straight line.

Using this numerical method to determine $q$ we find the
interesting results illustrated in Fig.\ref{fig3}. On the basis of the
earlier remarks making plausible that $q= 0$ at $ \mu = 2$, we
have been led to fit the numerical data with
\begin{equation}
q = (\mu - 2)^{\alpha},
\label{greatresult1}
\end{equation}
for $\mu \geq 2$
and
\begin{equation}
q = 1
\label{greatresult2}
\end{equation}
for $\mu \geq 3$.

We see from Fig.\ref{fig3} that the fitting function of
Eq. (\ref{greatresult1}) results in a satisfactory
agreement with the numerical result if we set
$\alpha \approx 0.13$. This means that
the critical value $q=0$
is reached with an infinite derivative, reinforcing
our conviction that $\mu = 2$ is a critical point of transition to
thermodynamics. The disorder in the region $2<\mu <3$ is partial, and
localized to the transition from one laminar string to another.
However, this is enough to generate a thermodynamic behavior.
The traditional wisdom would confine
thermodynamics to the region $\mu > 3$, which is where the
conventional central limit theorem applies. In a sense this analysis
shows that thermodynamics is possible also in the region where the
central limit theorem holds in the generalized form established by
L\'{e}vy \cite{LEVY,MONTROLL}. It is interesting to remark that
earlier research work\cite{RENATO,ELENA} has established
 that
the dynamical approach to diffusion, based on the
stationary assumption on the fluctuations responsible for diffusion,
is incompatible with the condition $\mu<2$. In this region a diffusion
process must rest on a continuous-time random walk method implying the
breakdown of the stationary assumption \cite{RENATO}.
 The region $ 2 \leq \mu < 3$ is compatible with stationary diffusion
 even if the diffusion process departs from ordinary Brownian diffusion
 and takes the shape of a L\'{e}vy process\cite{ALLEGRO}.
Therefore we conclude that
the stationary diffusion processes have the same regime of validity
as the non-extensive
thermodynamics of Tsallis. In fact, this paper
shows that the new perspective of Tsallis extends
the regime of validity of thermodynamics
to regions earlier imagined as being non thermodynamic,
in this case, to  $\mu < 3$. However, thermodynamics, even within
this new
perspective, cannot overcome the border $\mu = 2$. In other words,
it seems that the Tsallis thermodynamics
has the same regime of validity as the dynamic approach
 to diffusion, which is based on the assumption  that fluctuations
 are characterized by a stationary
 correlation function \cite{MANNELLA}.

 We have seen that the numerical calculations rests on both the
expedient of adopting $\tilde{\mu}$ rather than $\mu$ and that of
neglecting the border strings. The latter method is not only
an expedient to extend the regime of validity of our numerical
calculations. It reflects a property that probably deserves further
studies. In fact the border strings correspond to the peaks
that appear in the dynamical approach to the L\'{e}vy diffusion. As
discussed in Ref.\cite{ALLEGRO}, these peaks are a consequence
of the dynamic approach and the L\'{e}vy statistics are recovered
only in the time asymptotic limit. On the other hand, as pointed out
by the authors of\cite{ANNA}, a satisfactory agreement between the
entropic properties of trajectories and the general probabilistic
arguments of
\cite{probabilisticapproach1,probabilisticapproach2,probabilisticapproach3}
is obtained in the long-time regime, where the peak intensity tends to
vanish. This means, in other words, that the peaks seem to be
dynamic properties incompatible with the thermodynamic treatment,
 in accordance with the observation
made in this paper that the emergence of a $H_{q}(N)$ linearly
dependent on $N$ would be blurred by the presence of the two border
strings.

 We would be tempted to stress
that the detection of $q<1$ is expected
on the basis of the theoretical analysis
made by Lyra and Tsallis\cite{LYRA} on the dynamics of logistic map. These
authors show indeed that the generalization of the Pesin theorem to
the case of the logistic map implies $q<1$ at the chaos threshold. However, by the same
token we should conclude that these results disagree with those of Ref.
\cite{ANNA}, which rests, on the contrary, on a condition
physically much closer to that discussed in the present paper.
Actually, some caution must be
exerted in establishing a straight connection between the results of this paper
and the research work of \cite{ANNA,LYRA}, for the reasons pointed out earlier in this
paper. Here we are dealing with the ME that might imply
a so strong coarse graining as to lose a close relation
with the KST \cite{TPZ97}
 and consequently with the generalization
 of the important theorem of Pesin \cite{pesin}.
 This is made evident also by the
 fact that the correlated sequences
 are here generated by a stochastic approach, even if
 this turns out to be equivalent to
 the adoption of a nonlinear map \cite{nonlinearmap}. The statistical
 analysis in terms of the ME is insensitive of whether
 a nonlinear map or a stochastic approach is adopted. However,
 the peaks are dynamic properties that in all cases seem to be
 incompatible with the adoption of a merely
 entropic approach. This reinforces the need for carrying out
 the ME analysis by
 disregarding the
 border strings, as
 we propose in this paper.

 In summary, this paper sheds light into the breakdown of extensivity
 caused by time correlations. There are at the least two important
 sources of non extensivity: nonlocality in space \cite{ANTENEODO} and
nonlocality in
 time. For the spatial case, and up to now, there is a no clearcut
 connection in the literature between $q$ and the critical index
 characterizing spatial correlations (although there is a variety of
 strong indications). For the temporal case this manuscript
 establishes, for the first time, the analogous connection between $q$
 and $\mu$ (see Eq.(\ref{greatresult1}) and Eq.(\ref{greatresult2})).
 This is an interesting result and some efforts should be made to
 establish theoretically the critical exponent $\alpha$.

\newpage

\begin{figure}

\caption[]{The KST entropy as a function of $N$ in the completely
uncorrelated case (corresponding to $\mu = \infty$).
 In this case the KST entropy is expressed
by Eq. (\ref{nocorrelation}). For this reason the three curves have been
derived from
Eq. (\ref{nocorrelation}). The upper, middle and bottom lines refer to
$q=0.9$, $q=1$ and $q=1.1$, respectively.}
\label{fig1}
\end{figure}

\begin{figure}

\caption[]{The KST entropy as a function of $N$ with $\tilde\mu \cong \mu = 2.5 $. The three curves have
been obtained using the numerical treatment described in the text.
The upper (squares), middle (circles) and bottom (triangles) plots refer to
$q=0.82$, $q=0.89$ and $q=0.98$,
respectively.}
\label{fig2}
\end{figure}

\begin{figure}

\caption[]{ The entropic index $q$ versus $\tilde\mu$. See the text
for the
definition of $\tilde\mu$. The points with error bars are the result of
the numerical treatment described in the text, and the line
denotes the function $q=(\tilde\mu-2)^{\alpha}$ with $\alpha\approx 0.13$.}
\label{fig3}
\end{figure}


\begin{references}
\bibitem{nota} From the work of \cite{WANG}
we derive reliable prescriptions to evaluate
$h \equiv \lim_{N\rightarrow \infty} (H_{1}(N)/N)$. Adapting these
prescriptions to our notations, we see that this
limit is finite if $\mu>2$ and it vanishes for $\mu \leq 2$.
The non vanishing limit of $\mu$ is reached with a very
slow regression
process, with an inverse power law behavior. We are thus tempted
to interpret the region $ 2 < \mu \leq 3$ as one where
the coarse-grain induced randomness is not totally overcome by the
long-range correlations. We cannot rule out the possibility that the
entropic index $q$ undergoes a slow transition from $q<1$ to $q = 1$.
Rather, the conclusion of Wang \cite{WANG} can be made compatible with
the result of the numerical investigation of this paper by making
the plausible conjecture that the entropic index
$q < 1 $ is not permanent and undergoes a slow
regression to $q = 1$. The theoretical predictions of
\cite{WANG} suggests that the time evolution,
in the ``time'' N, of $q(N)$ towards
the asymptotic value $q( \infty ) = 1$ might take place with an
inverse power law behavior.
\end{references}
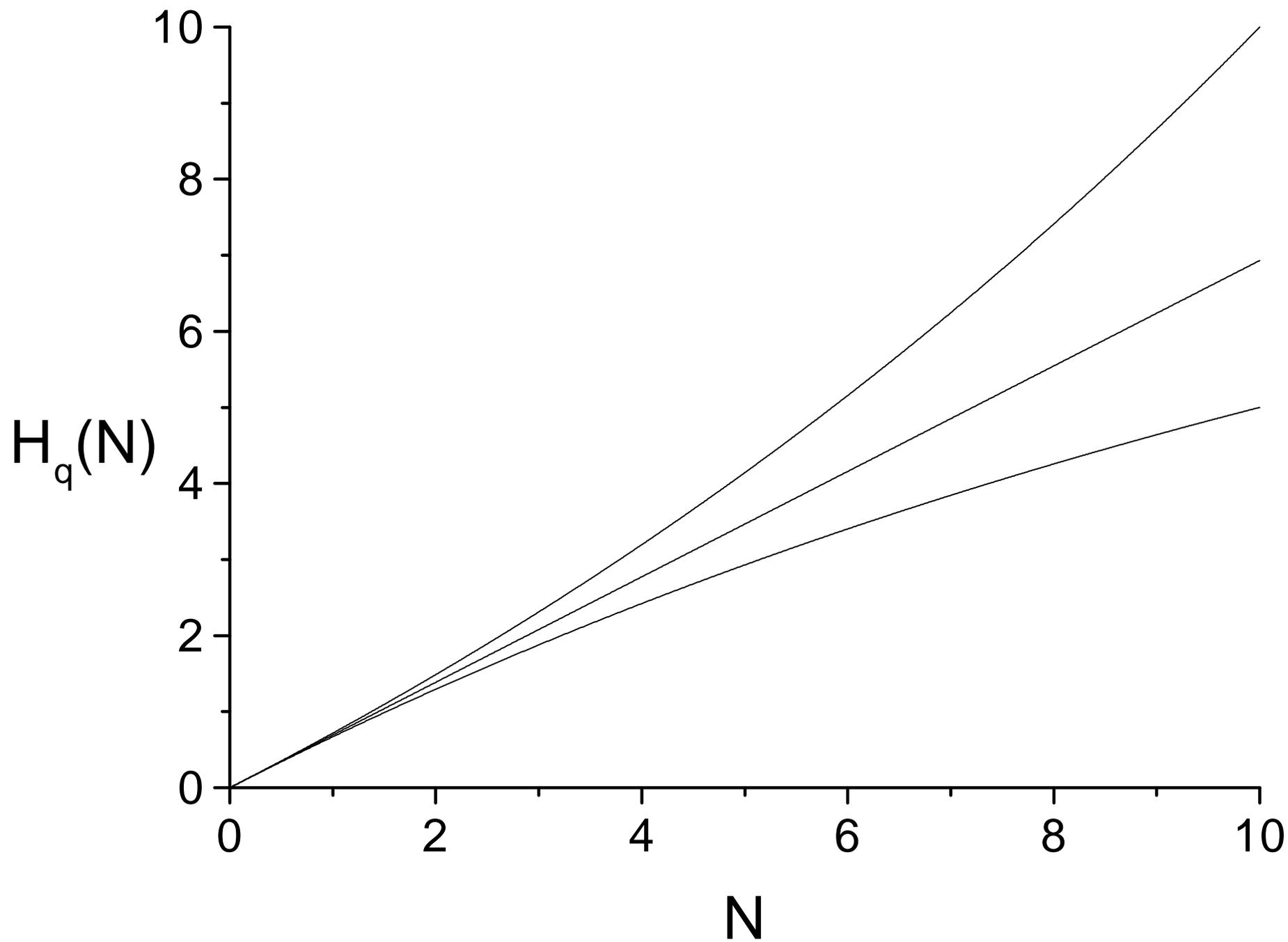
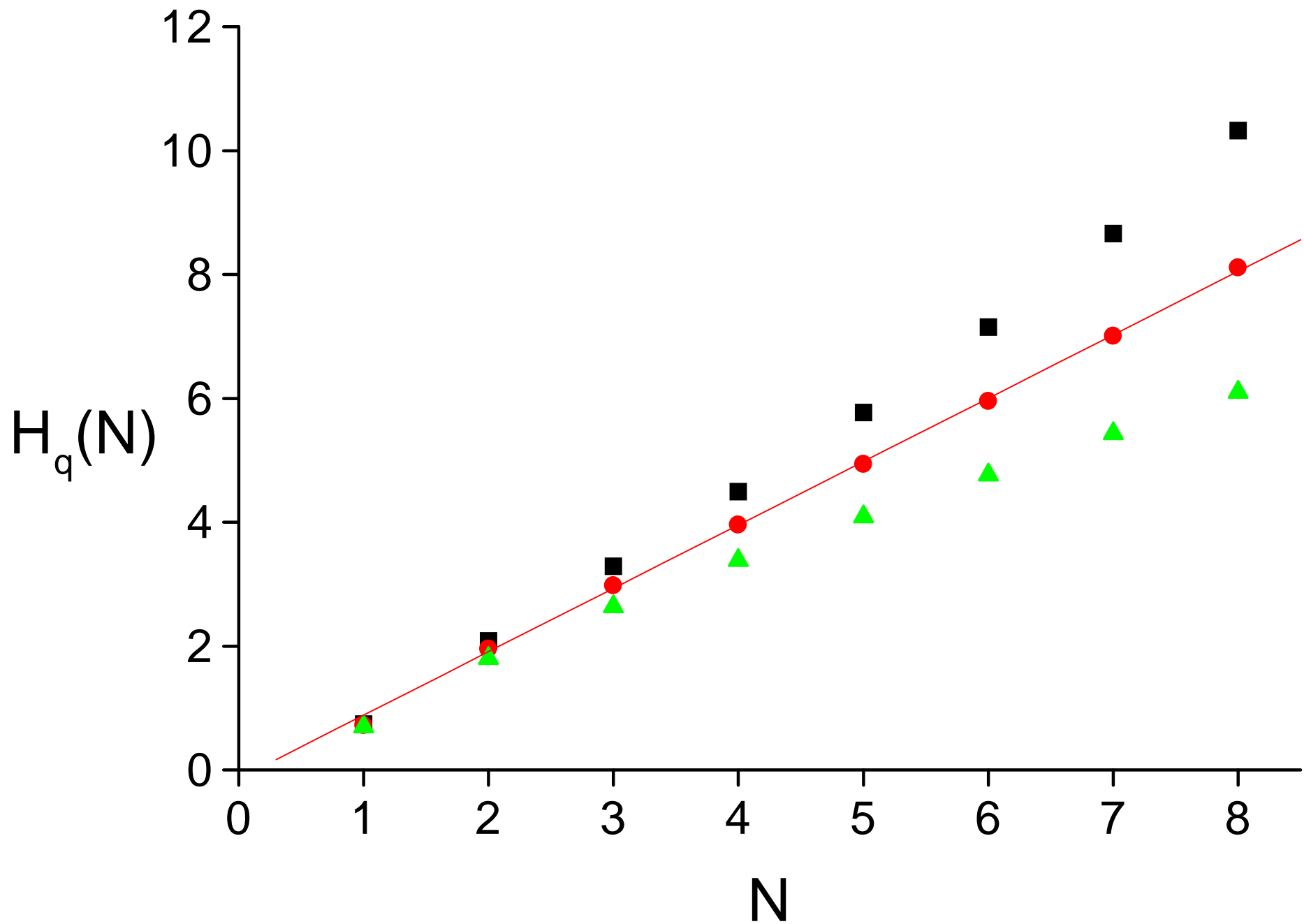
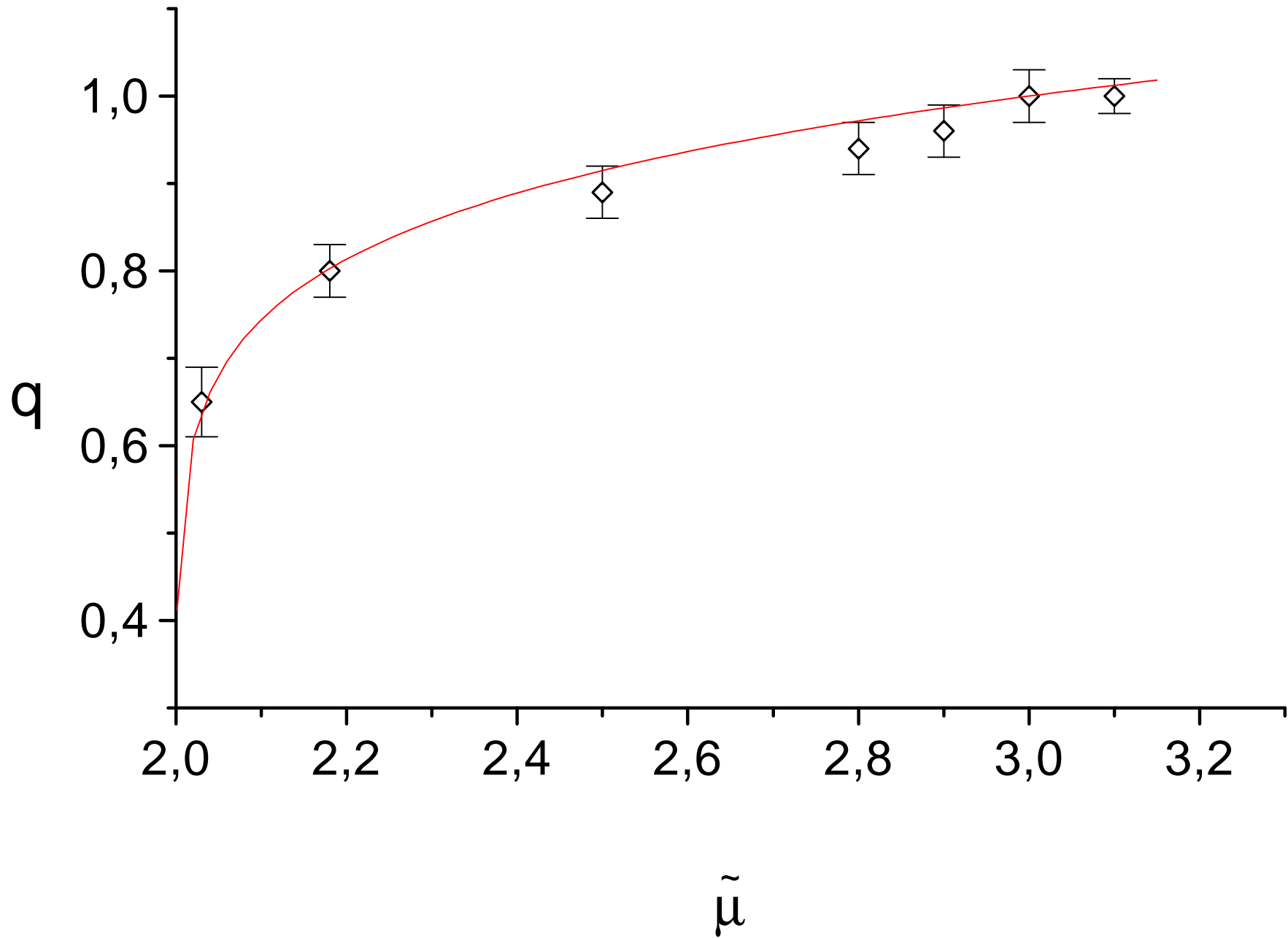
\end{document}